\documentclass[twocolumn,preprintnumbers,amsmath,amssymb]{revtex4}
\usepackage{graphicx}
\usepackage{dcolumn}% Align table columns on decimal point
\usepackage{bm}% bold math
\usepackage{float}
\usepackage{color}
\usepackage{float}
\begin{document}
%\begin{CJK*}{GBK}{song}%\begin{CJK*}{GBK}{song} %显示中文

\title{Controllable non-reciprocal transmission of single photon in M\"{o}bius
structure}% Force line breaks with \\

%\author{Hai-Yuan Zhu}$^{1,*}$
%\author{Jun-Jie Lin}$^{1,*}$
%$\footnote{These authors contributed equally to this work.}
%\author{Qing Ai}%
% \email{aiqing@bnu.edu,cn}
% \homepage{http://quanphys.bnu.edu.cn}
%\affiliation{%
%Department of Physics, Applied Optics Beijing Area Major Laboratory,
%Beijing Normal University, Beijing 100875, China
%}%
\author{Hai-Yuan Zhu,$^{*}$ Xin-Yuan Hu,$^{*}$  Jun-Jie Lin,$^{}$\footnote{These authors contributed equally to this work.}
Jia-Yi Wu, \\ Shuo Li, Yan-Xiang Wang ,
Fu-Guo Deng,
Na-Na Zhang$^{}$\footnote{201831140027@mail.bnu.edu.cn}}
%Qing Ai(艾清),$^{}$\footnote{aiqing@bnu.edu.cn\homepage{http://quanphys.bnu.edu.cn}}}
%

\address{Department of Physics, Applied Optics Beijing Area Major Laboratory,
Beijing Normal University, Beijing 100875, China\\}

%\affiliation{
%Second institution and/or address\\
%This line break forced% with \\
%}%

\date{\today}% It is always \today, today,
             %  but any date may be explicitly specified

\begin{abstract}
We propose a controllable non-reciprocal transmission model. The model consists of a M\"{o}bius ring, which is connected with two one-dimensional semi-infinite chains, and with a two-level atom located inside one of the cavities of the M\"{o}bius ring.
We use the method of Green function to study the transmittance of a single photon through the model. The results show that the non-reciprocal transmission can be achieved in this model and the two-level atom can behave as a quantum switch for the non-reciprocal transport of the single photon. This controllable non-reciprocal transmission model
may inspire new quantum non-reciprocal devices.
\end{abstract}

\maketitle

\section{\label{sec:Intro}Introduction}

According to Maxwell's electromagnetic equations, in most conditions,
the propagation of electromagnetic wave is reciprocal, that is, the transmission
of light from A to B is the same as that from B to A \cite{Griffiths1998}. However, because reciprocity greatly limits our processing of optical signals,  we need to find a way to break this symmetry. With
the development of physics, it has been recognized that quantum physics can provide a fertile ground where non-reciprocity concepts can be fruitfully exploited \cite{El-Ganainy2018} and metamaterials with negative refraction can be produced \cite{Zhao2020}.
Optical non-reciprocity refers to the
phenomenon that the propagation of light from one direction is different
from the propagation from the opposite direction. It has a great important role in protecting sensitive optical components \cite{Shoji2008}, constructing
quantum networks, and quantum signal processing \cite{jalas2013}.
Therefore, it is necessary to explore the optical devices which may achieve
non-reciprocity \cite{lei2011,dumlow1992,cirac1996,lira2012,horsley2013}. Traditional
non-reciprocal devices mainly include Faraday isolators which are based
on the magneto-optical effect \cite{tzuang2014} and some micro-resonators
which utilize the nonlinearity of materials \cite{bender2013}. However, the non-reciprocal
devices which are based on the magneto-optical effect usually have
large size because of the property of the materials. Moreover, non-linear devices
may be still reciprocal for some weak signals \cite{yu2015}. Therefore, people have
been exploring alternative strategies to realize non-reciprocal transmission,
for example, the use of atomic thermal motion \cite{zhang2018} and optical
power coupling \cite{manipatruni2009}, etc.
Here, we hope to find an effective, convenient and safe device with non-reciprocity.

In recent years, there has been a surge of interest in the theoretical and experimental study of the topological properties of quantum system  \cite{gravesen2005,hayashi2005,yakubo2003}, topologically-nontrivial structures has gradually attracted our attention. Furthermore, the molecule-based devices have motivated many investigations
because of their application prospect \cite{xu2008,liu2019,xu2019,guo2009}. Because of the unusual properties and potential applications \cite{fang2016}, M\"{o}bius structure has recently been under great focus both experimentally and theoretically, for example, the realization of negative refraction by using
M\"{o}bius structure \cite{ajami2003} and observing
various topological effects in M\"{o}bius structure \cite{zhao2009}.
Molecules with different structures can exhibit various novel physical properties.
Previous work have found that the transmission through M\"{o}bius ring
exhibits obvious differences from that of ordinary ring\textemdash the
transmission through the M\"{o}bius ring is suppressed in one direction \cite{zhao2009}. As M\"{o}bius structure has shown its development and application potential, this inspires us to use M\"{o}bius structure to realize non-reciprocity.

In this paper, we propose a controllable non-reciprocal transmission
model with single photon incidence. The model consists of a M\"{o}bius
ring connecting two semi-infinite chains. Both M\"{o}bius ring and semi-infinite chains are composed of a series of cavities. And a two-level atom
is embeded in one of the cavities of the M\"{o}bius ring. This atom behaves as a quantum switch that can control the non-reciprocal transport of the single photon.
Our research may inspire new quantum non-reciprocal devices.

The paper is organized as follows.
In Sec.~\ref{sec:Model}, we introduce
our model firstly, and then we consider two cases. One of the cases corresponds to that a two-level atom is added into one of the cavities of the M\"{o}bius ring. And the other one case corresponds to that there is no atom in the model. Then we calculate the transmittance of a single photon through the system by using Green function.
In Sec.~\ref{sec:Results},
we show and analyze the results of our calculation.
Finally, we discuss the prospect and conclusions in Sec.~\ref{sec:Conclusion}.

\section{\label{sec:Model}Model}

\begin{figure}
\includegraphics[width=9cm]{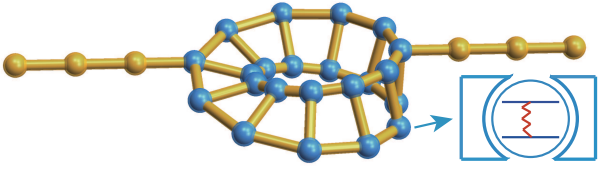}
\caption{Schematic diagram: the two ends of M\"{o}bius ring are connected with semi-infinite chains, and a two-level atom is added into one of the cavities of the M\"{o}bius ring.
Both M\"{o}bius ring and semi-infinite chains are composed of a series of cavities.
We mark that the connection cavity between the left (right) chain and the M\"{o}bius ring as $a_{L'}$ ($a_{R'}$), and the cavities of the upper (lower) ring of the M\"{o}bius ring as $a_{j}$ ($b_{j}$) ($j=0,1,...,N-1$). And we mark the cavities of the left and right chain as $c_{-l}$ and $c_{l}$ ($l=1,...,\infty$), respectively.
In addition, we label the cavity containing the two-level atom as $a_{n}$.
\label{fig:model}}
\end{figure}

In this paper, we consider a M\"{o}bius ring coupled with two semi-infinite
chains, and a two-level atom is embeded in one of the cavities of the M\"{o}bius ring as shown in
Fig.~\ref{fig:model}. The photon can transport through the chains
and the tunnel between the chains and the M\"{o}bius ring.
Let $a_{j}^{\dagger}$ $(a_{j})$
and $b_{j}^{\dagger}$ $(b_{j})$ be the creation(annihilation) operator
of the $j$th cavity of the upper and lower ring of M\"{o}bius ring, respectively.

\begin{figure}
\includegraphics[width=9cm]{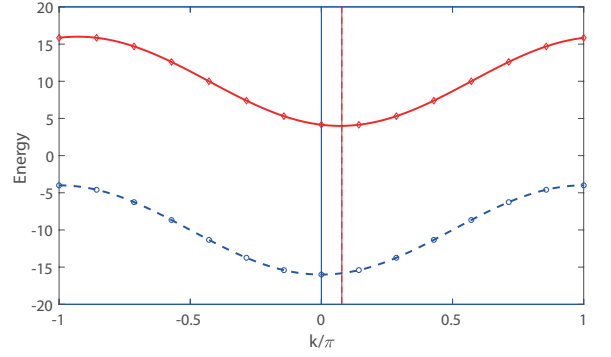}\caption{The energy spectrum of the M\"{o}bius ring ($N=7, V=10$ and $\xi=3$). The red solid line and the blue dashed line correspond to the upper and lower band of the M\"{o}bius ring, respectively. The red dashed line represents the symmetry axis of the upper band, which correspond to $k=\frac{\pi}{N}$. And the blue solid line represents the symmetry axis of the lower band, which correspond to $k=0$.\label{fig:Energy}}
\end{figure}

The Hamiltonian
of the M\"{o}bius ring reads \cite{zhao2009}
\begin{align}
H_{\text{M}}= & {\sum^{N-1}_{j=0}}[\varepsilon(a_{j}^{\text{\dag}}a_{j}+b_{j}^{\text{\dag}}b_{j})-V a_{j}^{\text{\dag}}b_{j} \nonumber\\
 & -\xi(a_{j}^{{\dag}}a_{j+1}+b_{j}^{{\dag}}b_{j+1})]+\textrm{h.c.},
\end{align}
where $V$ represents the coupling strength between $a_{j}$ and
$b_{j}$ sites, $\xi$ is the coupling strength between two adjacent cavities
and $\varepsilon$ describes the frequency of the cavity in the M\"{o}bius ring.
We assume that there are $2N$ cavities located at the upper and lower ring of M\"{o}bius ring. And the M\"{o}bius ring has boundary condition that $a_{N}=b_{0}$, $b_{N}=a_{0}$, which is different from ordinary rings.
This shows that the M\"{o}bius ring
do not obey the periodical boundary condition. By using the local
unitary transformation $c_{j_\uparrow}=\frac{1}{\sqrt{2}}(e^{i\frac{\varphi_{j}}{2}}a_{j}-e^{i\frac{\varphi_{j}}{2}}b_{j})$ and $c_{j\downarrow}=\frac{1}{\sqrt{2}}(a_{j}+b_{j})$, the Hamiltonian becomes
\begin{align}
H_{\text{M}}= & {\sum^{N-1}_{j=0}}[\varepsilon_{j}(e^{-i\varphi_{j}/2}c_{j\uparrow}^{\text{\dag}}c_{j\uparrow}+e^{i\varphi_{j}/2}c_{j\downarrow}^{\dagger}c_{j\downarrow})-V(c_{j\uparrow}^{\dagger}c_{j\downarrow} \nonumber\\
 & -c_{j\downarrow}^{\dagger}c_{j\uparrow})-\xi(e^{i\delta}c_{j\uparrow}^{\dagger}c_{j+1\uparrow}+c_{j\downarrow}^{\dagger}c_{j+1\downarrow}+\textrm{h.c.})],
\end{align}
where $\delta=\frac{\varphi_{j+1}}{2}-\frac{\varphi_{j}}{2}=\frac{\pi}{N}$ and $\varphi_{j}=j\frac{2\pi}{N}$.
The operator $c_{j\uparrow}$ $(c_{j\downarrow})$ satisfy the periodical
boundary condition. Then we use Fourier transformation $C_{k\uparrow}=\frac{1}{\sqrt{N}}\sum_{j}e^{ikj}c_{j\uparrow}$ and $C_{k\downarrow}=\frac{1}{\sqrt{N}}\sum_{j}e^{ikj}c_{j\downarrow}$ to diagonalize
the Hamiltonian and transform the Hamiltonian from real space to momentum space. By Fourier transformation, the Hamiltonian becomes
\begin{equation}
H_{\text{M}}=\sum_{k}(E_{k\uparrow}C_{k\uparrow}^{\text{\dag}}C_{k\uparrow}+E_{k\downarrow}C_{k\downarrow}^{\dagger}C_{k\downarrow}),
\end{equation}
where $E_{k\uparrow}=\varepsilon+V-2\xi\cos(k-\frac{\pi}{N}),$ $E_{k\downarrow}=\varepsilon-V-2\xi\cos(k)$, $k=j\delta$. In our calculation, we set $\varepsilon=0$.
According to the above calculation, we can draw the energy band diagram of the M\"{o}bius ring, as shown in Fig.~\ref{fig:Energy}. From Fig.~\ref{fig:Energy}, we can see that the lower energy band $E_{k\downarrow}$ is axially symmetric with respect to $k=0$. However, compared with the lower band, the symmetry axis of the upper band $E_{k\uparrow}$ is shifted.

The Hamiltonian of the semi-infinite chains \cite{zhou2008,zhou2013,you2003,you2005,chiorescu2004} is
\begin{align}
H_{\text{C}}= & \omega(\sum^{-1}_{l=-\infty}c_{l}^{\dag}c_{l}+\sum^{+\infty}_{l=1}
c_{l}^{\dag}c_{l})-\zeta[\sum^{-1}_{l=-\infty}(c_{l}^{\dag}c_{l+1}+\nonumber\\
 & +\sum^{+\infty}_{l=1}c_{l}^{\dag}c_{l+1})+\textrm{h.c.}],
\end{align}
where $\omega$ describes the frequency of the cavity in the chain, $\zeta$ is the coupling strength between two adjacent cavities.
 By using Fourier transformation $c_{k}=\frac{1}{\sqrt{M}}\sum_{l}e^{ikl}c_{l}$, the Hamiltonian can be rewritten as
\begin{equation}
H_{\text{C}}=\sum_{k}[\omega-2\zeta\cos(k)]c_{k}^{\dagger}c_{k},
\end{equation}
where $M$ is the total number of $l$.

The interaction Hamiltonian of the two semi-infinite chains and the M\"{o}bius ring is
\begin{align}
H_{\text{M}\text{C}}=\kappa(c_{-1}^{\dag}a_{L'}+c_{1}^{\dag}a_{R'}+\textrm{h.c.}),
\end{align}
where $\kappa$ represents the coupling strength between the two semi-infinite chains and the M\"{o}bius ring, the subscript $L'$ ($R'$) represents the connection cavity between the left (right) chain and the M\"{o}bius ring.
In the same way, by Fourier transformation we can also transform this Hamiltonian into the momentum space.

\begin{figure*}
\centering
\includegraphics[scale=0.42]{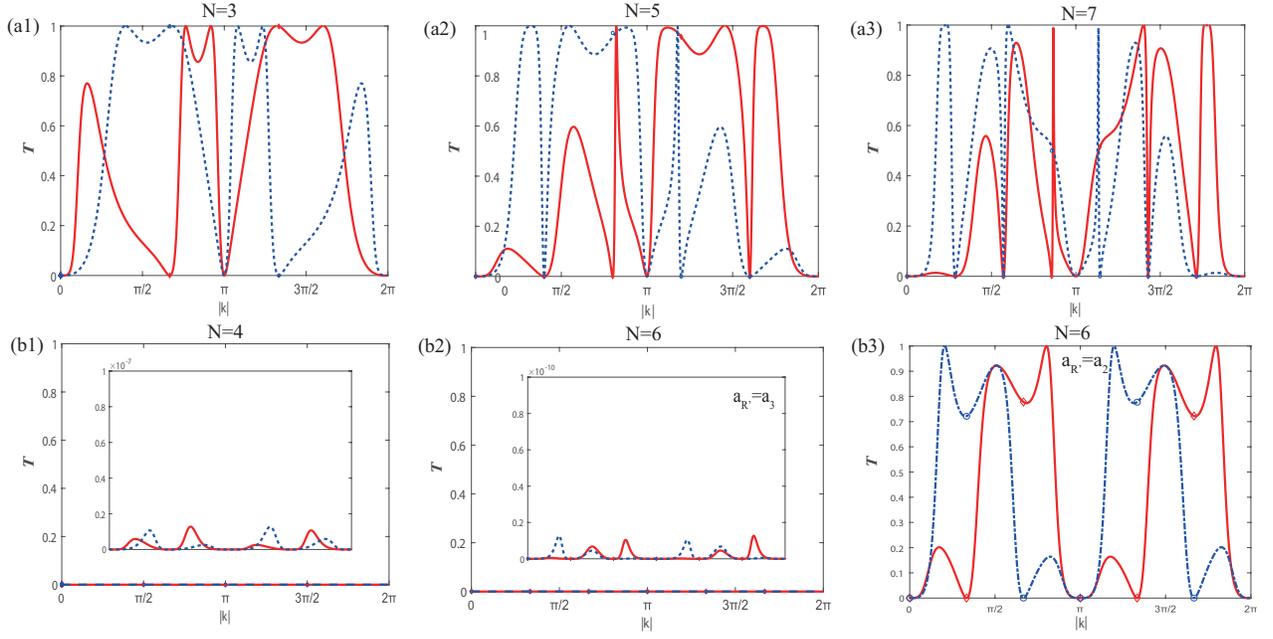}
\centering
\caption{The transmittance $T$ with respect to the absolute value of momentum $k$ of the incident photon. There are odd or even number of cavities in the M\"{o}bius ring, where the coupling strength $V=20$ and
$\xi=1$. And the photon is incident from the upper band. (a1), (a2), (a3) correspond to  $N=3,5,7$, respectively. In these three cases, the connection cavity between the left chain and the M\"{o}bius ring is $a_{L'}=a_{0}$, and that between the right chain and the M\"{o}bius ring is $a_{R'}=a_{\frac{N-1}{2}}$. (b1), (b2) correspond to $N=4,6$, respectively. And in these two cases, the connection cavity between the left chain and the M\"{o}bius ring is $a_{L'}=a_{0}$, and that between the right chain and the M\"{o}bius ring is $a_{R'}=a_{\frac{N}{2}}$. (b3) corresponds to $N=6$ and the connection cavity between the right chain and the M\"{o}bius ring change to $a_{R'}=a_{\frac{N}{2}-1}$. The red solid line represents the case of $k>0$ and the blue dashed line represents the case of $k<0$, where they represent two opposite photon incident directions.\label{fig3}}
\end{figure*}

When a two-level atom is added into the $n$th cavity of the M\"{o}bius ring, the total Hamiltonian of the system can be expressed as
\begin{equation}
H_{\text{S}}=H_{M}+H_{C}+H_{MC}+H_{A}+H_{AM},
\end{equation}
where $H_{A}$ is the Hamiltonian of the atom and $H_{AM}$ is the interaction Hamiltonian of the atom and the M\"{o}bius ring. Here
\begin{eqnarray}
H_{\text{A}}&=&\Omega_{A}d^{\text{\dag}}d, \\
H_{\text{A}\text{M}}&=&\gamma(a_{n}^{\text{\dag}}d+d^{\text{\dag}}a_{n}),
\end{eqnarray}
with $a_{n}=\frac{1}{\sqrt{2}}(e^{i\frac{\varphi_{n}}{2}}c_{j\uparrow}+c_{j\downarrow})=\sum_{k}\frac{1}{\sqrt{2}}e^{-ikn}(e^{i\frac{\varphi_{n}}{2}}C_{k\uparrow}+C_{k\downarrow})$,
where $\Omega_{A}$ describes the atom's frequency, and
$\gamma$ represents the coupling strength between the atom and the M\"{o}bius ring and the subscript $n$ represents that the atom is added into the $n$th cavity.
 After Fourier transformation we can obtain that
\begin{align}
H_{\text{A}\text{M}}= & \frac{\gamma}{\sqrt{2}}\sum_{k}[e^{ikn}(e^{-i\frac{\varphi_{n}}{2}}C_{k\uparrow}^{\dagger}+C_{k\downarrow}^{\dagger})d \nonumber\\
 & +d^{\text{\dag}}e^{-ikn}(e^{i\frac{\varphi_{n}}{2}}C_{k\uparrow}+C_{k\downarrow})].
\end{align}

We use the method of Green function to analyze the transmission of the single photon through the system \cite{datta2005}. The Green Function is
\begin{equation}
G=\frac{1}{(E+i0^{\text{+}})I-H-\varSigma_{L}-\varSigma_{R}},
\end{equation}
where $E$, $H$, $\varSigma_{L,R}$ correspond to the photon's incident frequency, the Hamiltonian of the M\"{o}bius ring and the self energies, respectively.
%In our system, self-energy is
%\begin{equation}
%\varSigma_{1}(a,a)=\varSigma_{2}(b,b)=-\kappa\exp(ik'),
%\end{equation}
%where $a,b$ depend on the connecting positions of the two chains and the M\"{o}bius ring. And the other elements of self-energy are zero.
By Fourier transformation, the self-energy in the momentum space can be expressed as
\begin{eqnarray}
\varSigma_{L(R)}&=&-\kappa e^{ik'}\sum_{k}\frac{1}{2N}(e^{-i\frac{\varphi_{L'(R')}}{2}}C_{k\uparrow}^{\dagger}+C_{k\downarrow}^{\dagger}) \\ \nonumber
&&\times(e^{i\frac{\varphi_{L'(R')}}{2}}C_{k\uparrow}+C_{k\downarrow}).
\end{eqnarray}
%where the subscript $L'$ ($R'$) represents the connection point between the right (left) chain and the M\"{o}bius ring.

Then the level broadenings can be expressed as
%($\varGamma_{1}=\varGamma_{2}=-2\text{Im}\varSigma_{1,2}$)
%\begin{equation}
%\varGamma_{1}=\varGamma_{2}=\sum_{k}\frac{1}{N}\kappa\sin k'(C_{k\uparrow}^{\dagger}C_{k\uparrow}+C_{k\downarrow}^{\dagger}C_{k\downarrow}).
%\end{equation}
\begin{eqnarray}
\varGamma_{1}&=&\text{i}\left[\varSigma-\varSigma^{\dagger}\right]_{L}.\\
\varGamma_{2}&=&\text{i}\left[\varSigma-\varSigma^{\dagger}\right]_{R}.
\end{eqnarray}
The transmission coefficient can be obtained by \cite{datta2005}
\begin{equation}
T=\mathrm{Tr}[\varGamma_{1}G\varGamma_{2}G^{\text{\dag}}]. \label{eq:T}
\end{equation}

\section{\label{sec:Results}Results}

We first consider the case that the photon is incident from
the upper band of the M\"{o}bius ring and the system without atoms.
According to Eq.~(\ref{eq:T}), we plot the photon's transmittance $T$ with respect to the momentum $k$, as shown in Fig.~\ref{fig3}.
Fig.~\ref{fig3} (a1), (a2), (a3) describe the change of transmittance T with $k$ when $N$ is taken as 3, 5, 7, respectively. In these figures, the connection cavity between the left (right) chain and the M\"{o}bius ring is $a_{L'}=a_{0}$ ($a_{R'}=a_{\frac{N-1}{2}}$).
Fig.~\ref{fig3} (b1), (b2) correspond to $N=4$ and $N=6$, respectively. In these two figures, the connection cavity between the right chain and the M\"{o}bius ring is $a_{R'}=a_{\frac{N}{2}}$.
Fig.~\ref{fig3} (b3) also corresponds to $N=6$ but change the connection cavity between the right chain and the M\"{o}bius ring to $a_{R'}=a_{\frac{N}{2}-1}$.
The red solid line and blue dashed line represent two opposite photon incident directions ($k>0$ or $k<0$).
From Fig.~\ref{fig3}(a) we can see that when $N$ is an odd number (3, 5, 7) and the photon is incident from the upper band that the non-reciprocity occurs.
This is because the upper band of the M\"{o}bius ring is asymmetric with respect
to $k=0$ as shown in Fig.~\ref{fig:Energy}.
It is also shows that when $N$ is an odd number, there is non-reciprocity at $k=\frac{N-1}{N}\pi$,
and with the increase of $N$, the effect of non-reciprocity decreases. This
is because when $N$ decreases, the shift of the symmetry axis of the upper energy band ($E_{k\uparrow}=\varepsilon+V-2\xi\cos(k-\frac{\pi}{N})$) will increase, and the asymmetry of M\"{o}bius ring will become more obvious.

Moreover, from Fig.~\ref{fig3} (b1), (b2) we can find that when $N$ is an even number and the connection cavity between the right chain and the M\"{o}bius ring is $a_{R'}=a_{\frac{N}{2}}$ , no matter what the value of $k$ is, there is no non-reciprocity.
The insets of Fig.~\ref{fig3} (b1), (b2) reduce the range of T values. From the insets, we can see that the difference between the two transmission curves can be seen only when $T$ is reduced to about the order of $10^{-10}$, and there is still no non-reciprocity at the point $k=j\frac{\pi}{N}$ with $j=0,1,...,N-1$. Based on Fig.~\ref{fig3} (b2), we change the connection cavity between the right chain and the M\"{o}bius ring to $a_{R'}=a_{\frac{N}{2}-1}$, and the curve of $T$ versus $k$ is shown in Fig.~\ref{fig3} (b3).
We can see that there is obvious non-reciprocity in this case.
We can explain this phenomenon from the structure of the model.
When $N$ is an even number and the connection cavity between the left (right) chain and the M\"{o}bius ring is $a_{L'}=a_{0}$ ($a_{R'}=a_{\frac{N}{2}}$), the structure of the model is symmetric, so there is no clear non-reciprocal behavior in this case, cf. Fig.~\ref{fig3} (b1), (b2).
When we change the connection cavity to $a_{R'}=a_{\frac{N}{2}-1}$, it is equivalent to breaking the structural symmetry of the model, so there is non-reciprocity, cf. Fig.~\ref{fig3} (b3).

\begin{figure}
\centering
\includegraphics[scale=0.6]{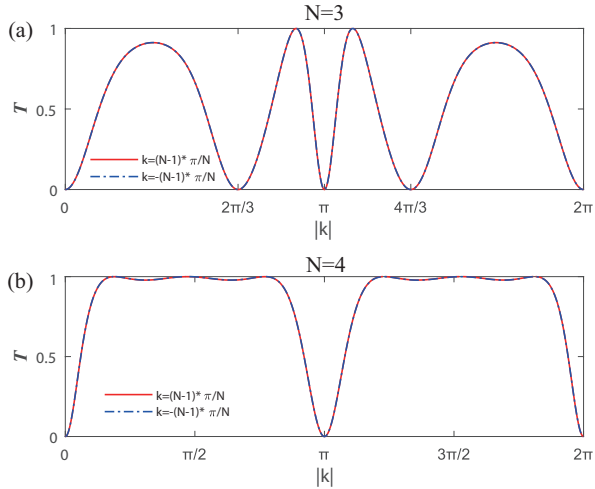}
\centering
\caption{The transmittance $T$ with respect to the absolute value of momentum $k$ of the incident photon. The coupling strength $V=20$ and
$\xi=1$. And the photon is incident from the lower band. In two figures, the connection cavity between the left chain and the M\"{o}bius ring is $a_{L'}=a_{0}$. (a) corresponds to $N=3$, and the connection cavity between the right chain and the M\"{o}bius ring is $a_{R'}=a_{\frac{N-1}{2}}$. (b) corresponds to $N=4$, and $a_{R'}=a_{\frac{N}{2}}$. The red solid line represents the case of $k>0$ and the blue dashed line represents the case of $k<0$, where they represent two opposite photon incident directions.\label{fig4}}
\end{figure}

Then we consider the case that the photon is incident from
the lower band of the M\"{o}bius ring and the system without adding atoms. Still keep the connection cavity of the left and right chains and the M\"{o}bius ring as $a_{L'}=a_{0}$, $a_{R'}=a_{\frac{N}{2}}$ or $a_{R'}=a_{\frac{N-1}{2}}$. The calculation results are shown in Fig.~\ref{fig4}. We can find that when photon is incident from the lower band, no matter weather $N$ is an odd or even number, there is no non-reciprocity under different $k$ values. This is mainly because the lower band of the M\"{o}bius ring is symmetric with respect
to $k=0$ as shown in Fig.~\ref{fig:Energy}.

\begin{figure*}
\includegraphics[scale=0.42]{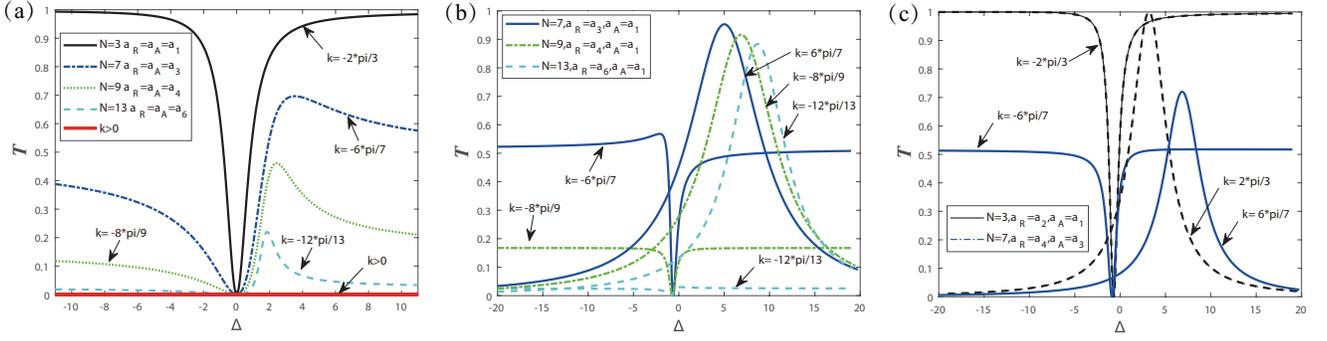}
\caption{The transmittance $T$ with respect to $\Delta$, where $\Delta$ represents the detuning between the frequency of incident photon and the frequency of the atom. Consider adding a two-level
atom into one of the cavities of the M\"{o}bius ring, where $V=20$, $\xi=1$ and $\kappa=3$. The photon is incident from the upper band. (a) $N=3,7,9,13$, the atom is added into the $a_{\frac{N-1}{2}}$ cavity and the right chain is connected to the $a_{\frac{N-1}{2}}$ cavity ($a_{n}=a_{\frac{N-1}{2}}, a_{R'}=a_{\frac{N-1}{2}}$). For different $N$, we take the momentum $k$ of the incident photon as $k=\pm\frac{(N-1)\pi}{N}$. The red dashed line represents the case of $k=\frac{(N-1)\pi}{N}$. (b) correspond to $N=7,9,13$ and $a_{n}=a_{1}, a_{R'}=a_{\frac{N-1}{2}}$.
(c) correspond to $N=3,7$ and $a_{n}=a_{\frac{N-1}{2}}, a_{R'}=a_{\frac{N+1}{2}}$.
\label{fig5}}
\end{figure*}
Then, we add a two-level atom into one of the
cavities of the M\"{o}bius ring. We plot the transmittance $T$ of single photon with respect to $\Delta$, where $\Delta$ represents the detuning between the frequency of incident photon and the frequency of the atom.
In Fig.~\ref{fig5}, we consider the case that the photon is incident from the upper band of the M\"{o}bius ring and $N$ is taken as an odd number.
For different $N$, we take the momentum of the incident photon as $k=\pm\frac{(N-1)\pi}{N}$.
The positive and negative signs represent the different incident directions of the photon.
Fig.~\ref{fig5}(a) corresponds to the situation that the atom is added into the connection cavity between the right chain and the M\"{o}bius ring, that is, $a_{n}=a_{R'}=a_{\frac{N-1}{2}}$.
Fig.~\ref{fig5}(b) corresponds to the case that the connection cavity between the right chain and the M\"{o}bius ring unchanged, which is $a_{R'}=a_{\frac{N-1}{2}}$, but the atom is added into the cavity $a_{n}=a_{1}$ .
Fig.~\ref{fig5}(c) corresponds to that the cavity of the atom is added is unchanged, which is $a_{R'}=a_{\frac{N-1}{2}}$, but the connection cavity between the right chain and the M\"{o}bius ring change to $a_{R'}=a_{\frac{N+1}{2}}$.
From Fig.~\ref{fig5}(a) we can see that for different $N$, except for $\Delta=0$, there is non-reciprocity in other positions, and the smaller $N$ is, the stronger the non-reciprocity is.
The red line in the figure corresponds to the cases that $k=\frac{(N-1)\pi}{N}$ ($N=3,7,9,13$). This line shows that for different $N$, when the momentum of the incident photon is $k=\frac{(N-1)\pi}{N}$, the transmittance $T$ of the photon is zero at any $\Delta$.
Comparing Fig.~\ref{fig5}(a) with Fig.~\ref{fig5}(b), we can see that changing the position of the atom will affect the non-reciprocity of the system.
When the position of the atom is changed from $a_{n}=a_{\frac{N-1}{2}}$ to $a_{n}=a_{1}$, we can see that for $N=7,9,13$, the transmission corresponding to $k=\frac{(N-1)\pi}{N}$ is no longer zero under different $\Delta$ values, and the non-reciprocity also appears at the point of $\Delta=0$.
Similarly, comparing Fig.~\ref{fig5}(a) with Fig.~\ref{fig5}(c), we can see that changing the connection cavity between the right chain and the M\"{o}bius ring will also affect the non-reciprocity of the system.
Fig.~\ref{fig5}(c) also shows that for $N=3,7$, the transmission corresponding to $k=\frac{(N-1)\pi}{N}$ is no longer zero under different $\Delta$ values, and the non-reciprocity also appears at the point of $\Delta=0$.
According to Fig.~\ref{fig5}, we can conclude that adding a two-level atom into one of the cavities of the M\"{o}bius ring will affect the non-reciprocity of the system. Therefore, the atom can be used as a quantum switch to control the non-reciprocity of the system. Moreover, different positions of the atom will bring different results.
In addition, the different connection cavities between the right chain and the M\"{o}bius ring will also affect the non-reciprocity of the system.
Therefore, we can adjust the non-reciprocity of the system from these three aspects.

\begin{figure}
\includegraphics[scale=0.4]{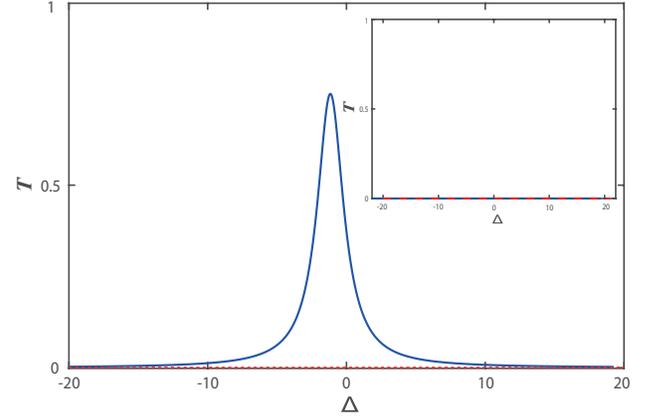}
\caption{The transmittance $T$ with respect to $\Delta$ under the condition that a two-level
atom is added into one of the cavities of the M\"{o}bius ring, where $V=20$, $\xi=1$, $\kappa=3$ and $N=6$. The photon is incident from the upper band, and the atom is added into the $a_2$ cavity. The inset represents that the atom is added into the $a_3$ cavity. The red solid line represents the case of $k=\frac{4\pi}{6}$ and the blue dashed line represents the case of $k=-\frac{4\pi}{6}$. The connection cavity between the left (right) chain and the M\"{o}bius ring is $a_{L'}=a_{0}$ ($a_{R'}=a_{3}$). \label{fig6}}
\end{figure}

Next, we consider the case that $N$ is an even number. We can find that when $N$ is an even number and the symmetry of the model structure is kept, i.e. $a_{L'}=a_{0},a_{R'}=a_{\frac{N}{2}}$, adding a two-level atom into one of the cavities of the M\"{o}bius ring leads to the non-reciprocity of the system, cf. Fig.~\ref{fig6}.
However, when the atom is added into the $a_{n}=a_{\frac{N}{2}}$ cavity, there is no non-reciprocity occurs, as shown in the inset of Fig.~\ref{fig6}.
We think that the reason for this phenomenon may be that the symmetry of the structure will not be destroyed when the atom is added into the cavity of $a_{n}=a_{\frac{N}{2}}$.
Compared with Fig.~\ref{fig4}, we can conclude that when $N$ is an even number, the non-reciprocity of the system can be realized by adding a two-level atom into one of the cavities of the M\"{o}bius ring. Therefore, when $N$ is an even number, the atom can still be used as a quantum switch to control the non-reciprocity of the system.

\section{\label{sec:Conclusion}Conclusion}
In this paper, we explore the method of realizing non-reciprocity in the system based
on M\"{o}bius structure.
We first analyze the transmission of a single photon through the system without atoms.
We find that when the photon is incident from the upper band of the M\"{o}bius
ring and the number of cavities in the M\"{o}bius
ring is odd, the non-reciprocal transmission can be realized, but
not when the number of cavities is even.
And we find that with the increase of $N$, the effect of non-reciprocity decreases, cf. Fig.~\ref{fig3} (a).
This is because when $N$ decreases, the shift of the symmetry axis of the upper energy band $E_{k\uparrow}$ will increase, and the asymmetry of the M\"{o}bius ring will become more obvious. Then, we consider the case that the number of cavities in the M\"{o}bius
ring is even. By calculation, we find that when $N$ is an even number and the connection cavity between the right chain and the M\"{o}bius ring is changed to $a_{R'}\neq a_{\frac{N}{2}}$, the non-reciprocity occurs, cf. Fig.~\ref{fig3} (b3). The reason for this phenomenon can be explained from the structure of the model, that is, the change of the connection cavity between the right chain and the M\"{o}bius ring destroys the structural symmetry of the model.
However, because the lower band of the M\"{o}bius ring is symmetric with respect to the momentum
$k=0$, whether $N$ is an even or odd number, the non-reciprocal transmission can not be found when the photon is incident from the lower band, cf. Fig.~\ref{fig4}.
We then consider the case that a two-level atom is added into one of the
cavities of the M\"{o}bius ring. By analyzing the photon's transmittance $T$ with respect to detuning between the frequency of the incident photon and the frequency of the atom, we find that adding a two-level atom into one of the cavities of the M\"{o}bius ring will affect the non-reciprocity of the system, and different adding positions of the atom will bring different results.
Therefore, the atom can act as a quantum switch to control the non-reciprocity of the system.
In addition, the different connection sites between the right chain and the M\"{o}bius ring will also affect the non-reciprocity of the system.
We also show that when $N$ is an even number and the symmetry of the model structure is kept, i.e. $a_{L'}=a_{0}, a_{R'}=a_{\frac{N}{2}}$, adding a two-level atom into one of the cavities of M\"{o}bius ring ($a_{n}\neq a_{\frac{N}{2}}$) leads to the non-reciprocity of the system.
Therefore, we can conclude that no matter weather $N$ is an odd or even number, the atom can be used as a quantum switch to control the non-reciprocity of the system.
This controllable non-reciprocal transmission model
may inspire new quantum non-reciprocal devices.

\begin{acknowledgments}
This work is supported by the National Natural Science Foundation of China under Grant Nos.~11674033,~11474026,~11505007, and Beijing Natural Science Foundation under Grant No.~1202017.
\end{acknowledgments}

\color{black}

%\end{CJK*}

\end{document}